\documentclass[twocolumn,showpacs,floats,floatfix,superscriptaddress,aps,pra]{revtex4-1}
\usepackage{amsfonts}
\usepackage{amssymb}
\usepackage{amsmath}
\usepackage{calc}
\usepackage{graphicx}
\usepackage{bm}

\usepackage[normalem]{ulem}
\usepackage{amsmath,amssymb}
\usepackage{multirow}
\usepackage{xcolor,soul}
\usepackage{srcltx}
\usepackage{hyperref,graphicx}

\begin{document}

\author{Stefano Longhi$^{*}$} 
\affiliation{Dipartimento di Fisica, Politecnico di Milano and Istituto di Fotonica e Nanotecnologie del Consiglio Nazionale delle Ricerche, Piazza L. da Vinci 32, I-20133 Milano, Italy}
\email{stefano.longhi@polimi.it}

\title{Probing non-Hermitian Skin Effect and non-Bloch Phase Transitions}
  \normalsize


%
\bigskip
\begin{abstract}
\noindent  
In non-Hermitian crystals showing the non-Hermitian skin effect,
ordinary Bloch band theory and Bloch topological invariants fail to correctly predict energy spectra, topological boundary states, and symmetry breaking phase transitions in systems with open boundaries. Recently, it has been shown that a correct description requires to extend Bloch band theory into complex plane.  A still open question is whether non-Hermitian skin effect and non-Bloch symmetry-breaking phase transitions can be probed by real-space wave dynamics far from edges, which is entirely governed by ordinary Bloch bands. Here it is shown that the Lyapunov exponent in the long-time behavior of bulk wave dynamics can reveal rather generally non-Bloch symmetry breaking phase transitions and the existence of the non-Hermitian skin effect.

\end{abstract}

\maketitle

\section{Introduction}
Bloch band theory describes energy spectra and single electronic bulk states in crystals with either periodic boundary conditions (PBC) or open boundary conditions (OBC).
Remarkably, Bloch bulk invariants can be introduced to classify topological bands and to predict the appearance of topological edge states in crystals with OBC (bulk-boundary correspondence) \cite{R1,R2,R3}.
Such major results are challenged when trying to apply Bloch band theory to non-Hermitian systems. Topological characterization of non-Hermitian models is currently a hot area of research
(see \cite{R3bis,R4,R5,R6,R7,R8,R9,R9bis,R9tris,R9quatris,RLonghi} and references therein). Among the most relevant features observed in non-Hermitian systems, one should mention the strong sensitivity of the energy spectra on boundary conditions \cite{R6,R10,R11,BB1,BB2,R12,R13,R14}, the non-Hermitian skin effect (NHSE) \cite{R14,R6,R8,BB1,BB2,R13,R13bis,R15,Ruff1}, i.e. the exponential
localization of continuum-spectrum eigenstates to the edges, and the failure of the bulk-boundary correspondence based on Bloch band topological invariants \cite{R3bis,BB1,BB2,BB3,BB4,BB5,BB5bis,BB5quatris,BB6,BB7,BB8,R16,R16bis,R16tris,Ruff1,Ruff2,Ruff3}. Recently, several attempts have been suggested to restore the bulk-boundary correspondence, such as those based on the biorthogonal bulk-boundary correspondence \cite{BB1}, the non-Bloch bulk topological invariants \cite{BB2,BB3,BB4,BB5,BB5bis,BB5quatris,R16tris}, the singular value decomposition \cite{BB6}, and the Green functions \cite{BB7,BB8}. A major consequence of the NHSE is that the bulk bands of the OBC system are considerably different from those of the PBC
system. While the latter are defined by ordinary Bloch band theory, the former are non-Bloch bands that require the quasi-momentum to become complex and to vary on a generalized Brillouin
zone \cite{BB2,BB3,BB4}. The usefulness of non-Bloch band theory in non-Hermitian systems is demonstrated by restoration of (non-Bloch) bulk-boundary correspondence \cite{BB2,BB3,BB4,BB5,BB5bis,BB5quatris}
and in the study of non-Hermitian wave scattering and domain walls \cite{R16tris,DW}. 
Another major consequence of the NSHE is that distinct bulk symmetry breaking phase transitions are observed when considering Bloch and non-Bloch bands, i.e. systems with PBC and OBC. For example, for certain non-Hermitian extensions of the Su-Schrieffer-Heeger (SSH) model \cite{R17}, the bulk eigenenergies in the case of OBC are entirely real over a wide range of system parameters as a consequence of pseudo-Hermiticity, while they are complex for PBC \cite{R3bis,BB2,BB5bis,R13}. A similar result was found in one-band systems where the NHSE arises from an imaginary gauge field \cite{R6,R13bis,R18}.\\ 
Bulk dynamics in real space, such as in quantum walk experiments, is a powerful tool to provide useful information about topological invariants and edge states in synthetic topological matter \cite{S1,S2,S3,S4,S5,S6,S7,S8,S9}. A natural question then arises: can NHSE and symmetry breaking phase transitions of non-Bloch bands be probed looking at the bulk wave dynamics? At first sight, one would expect such a program to fail because the bulk motion of  a wave packet, far from edges, is entirely determined by the structure of Bloch bands. How could such a wave packet feel boundary effects and non-Bloch band features, given that it is the superposition of ordinary bulk (extended) Bloch states?  Contrary to such a wisdom, in this work it is shown that  both non-Bloch symmetry-breaking phase transitions and the NHSE can be probed looking at the time behavior of wave dynamics in the bulk. Indeed, the long-time behavior of a wave packet in a system with PBC is established by the turning points of non-Bloch bands, which can reveal both NHSE and non-Bloch symmetry-breaking phase transitions.\\ 
The paper is organized as follows. In Sec. II we introduce the general two-band non-Hermitian model in real space and Bloch space under PBC and OBC, with special focusing onto four non-Hermitian extensions of the SSH model. A sufficient criterion for the existence of the NHSE, based on saddle points of the band dispersion curves, is presented in Sec.III. The interplay between NHSE and the Lyapunov exponent in real-space dynamics is highlighted in Sec.IV, whereas bulk probing of non-Bloch phase transitions is discussed in Sec.V.  Finally, the main conclusions and future outlook are given in Sec.VI.   
\section{Two-band non-Hermitian models}
\subsection{Real space Hamiltonian, Bloch Hamiltonian, and boundary conditions}
We consider a one-dimensional (or a quasi one-dimensional) tight-binding lattice with two sites per unit cell. Indicating by $a_n$ and $b_n$ the occupation amplitudes in the two sublattices A and B at the $n$-th unit cell,
the real-space dynamics is governed by the coupled equations
\begin{eqnarray}
i \frac{da_n}{dt} & = \sum_l \rho_{n-l} a_l+ \sum_l \theta_{n-l} b_l \\
i \frac{db_n}{dt} & = \sum_l  \varphi_{n-l} a_l- \sum_l \rho_{n-l} b_l 
\end{eqnarray}
where $\rho_n$ ($n \neq 0$) are the intra-sublattice hopping amplitudes, $\pm \rho_0$ are the on-site potentials, and $\theta_n$, $\varphi_n$ are the (generally asymmetric) inter-sublattice hopping amplitudes. A Hermitian lattice corresponds to $\rho_{-n}=\rho_n^*$ and $\theta_{-n}=\varphi_{n}^*$. Owing to the NHSE, the energy spectrum and corresponding eigenfunctions are strongly dependent on the boundary conditions. Here we consider either PBC or OBC.\\ 
\\
{ {\it Lattice with PBC}.} For an infinitely-extended lattice or for a lattice with a finite number of unit cells and with PBC, one can set
\begin{equation}
\left(
\begin{array}{c}
a_n \\
b_n
\end{array}
\right)=\left(
\begin{array}{c}
A \\
B
\end{array}
\right) \exp(ikn-iEt)
\end{equation}
where $k$ is the Bloch wave number and $E=E(k)$ is the dispersion curve of the Bloch band. The wave number $k$ varies in the first Brillouin zone $-\pi \leq k < \pi$, and eventually it is quantized owing to the PBC.
Substitution of Eq.(3) into Eqs.(1) and (2) yields
\begin{equation}
E \left(
\begin{array}{c}
A \\
B
\end{array}
\right)
=H(k) \left(
\begin{array}{c}
A \\
B
\end{array}
\right)
\end{equation}
where  $H(k)$ is the $2 \times 2$ Bloch Hamiltonian in momentum space
\begin{eqnarray}
H(k) & = & \left(
\begin{array}{cc}
d_z(k) & d_x(k)-id_y(k)  \\
d_x(k)+id_y(k) & -d_z(k)
\end{array}
\right) \nonumber \\
& = & \sigma_xd_x(k)+\sigma_y d_y(k)+ \sigma_z d_z(k),
\end{eqnarray}
$\sigma_{x,y,z}$ are the Pauli matrices,
and where we have set
\begin{eqnarray}
d_x(k) & \equiv & \frac{1}{2} \sum_n (\theta_n+\varphi_n) \exp(-ikn)  \\ 
d_y(k) & \equiv & \frac{1}{2i} \sum_n (\varphi_n-\theta_n) \exp(-ikn) \\
d_z(k) & \equiv & \sum_n \rho_n \exp(-ikn).
\end{eqnarray}
Since in systems with PBC $k$ spans the first Brillouin zone, $\beta \equiv \exp(ik)$ varies on the unit circle $C_{\beta}$ in complex plane, i.e. $|\beta|=1$. The energy spectrum shows chiral symmetry with the dispersion curves of the two bands given by 
\begin{equation}
E_{PBC}=E_{\pm}(k)= \pm \sqrt{Q(\beta)}, 
\end{equation}
where
\begin{equation}
Q(k) \equiv d_x^2(k)+d_y^2(k)+d_z^2(k).
\end{equation}
We assume that the two bands are separable, i.e. $Q(\beta) \neq 0$ as $\beta=\exp(ik)$ varies on the unit circle $C_{\beta}$, corresponding to the absence of exceptional points (EPs).
Rather generally, $Q(\beta)$ is given by a sum of powers of $\beta$, i.e. 
\begin{equation}
Q(\beta)=\sum_n \sigma_n \beta^n= \sum_n \sigma_n \exp(ikn)
\end{equation}
where the number of terms in the sum is finite for limited long-range interactions. This readily follows from Eqs.(6), (7), (8) and (10), with the Fourier coefficients $\sigma_n$ of $Q(k)$ determined from those of $d_x(k)$, $d_y(k)$ and $d_z(k)$. Assuming (as it is physically reasonable) that the long-range hopping amplitudes vanish as $|l-n|$ is large, one can assume $\sigma_n=0$ for large enough $| n |$.\\
Finally, it is worth mentioning that the properties of the two-band Hamiltonian $H(k)$ in momentum space can be retrieved from the one of a single-band system with Hamiltonian $Q(k)$. In fact, from the eigenvalue equation (4) one has
\begin{equation}
E^2 \left(
\begin{array}{c}
A \\
B
\end{array}
\right)=H^2(k) 
\left(
\begin{array}{c}
A \\
B
\end{array}
\right)
\end{equation}
with $H^2(k)$ diagonal and given by
\begin{equation}
H^2(k)= \left(
\begin{array}{cc}
Q(k) & 0 \\
0 & Q(k)
\end{array}
\right).
\end{equation}
$Q(k)$ can be viewed as the Bloch Hamiltonian of a one-dimensional lattice with one site per unit cell and with hopping amplitudes $\sigma_n$, according to Eq.(11).\\
\\
{\it Lattice with OBC.} For a lattice comprising $N$ unit cells with OBC, let us set  $\psi_A=(a_1,a_2,...,a_N)^T$ and $\psi_B=(b_1,b_2,...,b_N)^T$. The coupled-equations  (1) and (2) can be cast in the compact form
\begin{equation}
i \frac{d}{dt}
\left(
\begin{array}{c}
\psi_A \\
\psi_B
\end{array}
\right)
= \left(\begin{array}{c|c} \mathcal{A} & \mathcal{B}_1 \\\hline \mathcal{B}_2 & -\mathcal{A} \end{array}\right) 
\left(
\begin{array}{c}
\psi_A \\
\psi_B
\end{array}
\right)
\end{equation}
where the elements of the $N \times N$ matrices $\mathcal{A}$, $\mathcal{B}_1$ and $\mathcal{B}_2$ are given by
\begin{equation}
\mathcal{A}_{n,l}=\rho_{n-l} \; , (\mathcal{B}_1)_{n,l}=\theta_{n-l} \; , \; (\mathcal{B}_2)_{n,l}= \varphi_{n-l} .
\end{equation}
\\
After setting $ \psi_A= a \exp(-iEt)$, $\psi_B= b \exp(-iEt)$, the energy spectrum $E$ for the system with OBC is obtained from the eigenvalue problem
\begin{equation}
E \left(
\begin{array}{c}
a \\
b
\end{array}
\right)
= \mathcal{H}
\left(
\begin{array}{c}
a \\
b
\end{array}
\right)
\end{equation}
where we have set
\begin{equation}
\mathcal{H} \equiv \left(\begin{array}{c|c} \mathcal{A} & \mathcal{B}_1 \\\hline \mathcal{B}_2 & -\mathcal{A} \end{array}\right).
\end{equation}
As shown in several recent works, the bulk OBC spectrum $E_{OBC}$, i.e. the spectrum of the matrix $\mathcal{H}$  in the large $N$ limit (disregarding possible isolated eigenvalues related to boundary states), can be obtained from Eq.(9) provided that $\beta$ varies on a generalized Brillouin zone $\tilde{C}_{\beta}$ which deviates from the unit circle $C_{\beta}$ \cite{BB4,BB5}. The generalized Brillouin zone is basically defined by the locus of $\beta$ such that on $\tilde{C}_{\beta}$ one can always find two points $\beta_{1}$ and $\beta_2$ with $|\beta_1|=|\beta_2|$ and $Q(\beta_1)=Q(\beta_2)$ (for a more precise definition of the generalized Brillouin zone see \cite{BB4,BB5}; see also \cite{R14}). To study the bulk energy spectrum with OBC, i.e. to determine the extended Brillouin zone $\tilde{C}_{\beta}$, let us consider the large $N$ limit. In this limit, one can assume $\mathcal{A} \mathcal{B}_{1,2} \simeq \mathcal{B}_{1,2} \mathcal{A}$ and $\mathcal{B}_1 \mathcal{B}_2 \simeq \mathcal{B}_2 \mathcal{B}_1$. This is because the elements of the commutator matrices $ [  \mathcal{A}, \mathcal{B}_{1,2} ] $ and $ [ \mathcal{B}_1, \mathcal{B}_2] $  are non-vanishing just near the edges and thus they do not influence the asymptotic behavior $(a,b) \propto \beta^n$ of bulk states, which determines $\tilde{C}_{\beta}$.  From Eq.(16) one then obtains
 \begin{equation}
 E^2 \left(
\begin{array}{c}
a \\
b
\end{array}
\right) = \mathcal{H}^2 \left(
\begin{array}{c}
a \\
b
\end{array}
\right).
 \end{equation}
 with
 \begin{equation}
 \mathcal{H}^2  \simeq \left(\begin{array}{c|c} \mathcal{A}^2+\mathcal{B}_1 \mathcal{B}_2 & 0 \\\hline 0 & \mathcal{A}^2+\mathcal{B}_1 \mathcal{B}_2  \end{array}\right)
 \end{equation}
 The bulk energy spectrum of the system with OBC is thus given by $E_{\pm}= \pm \sqrt{\Lambda}$, where $\Lambda$ are the eigenvalues of the $N \times N$ matrix $\mathcal{H}_0$ defined by
 \begin{equation}
 \mathcal{H}_0 \equiv \mathcal{A}^2+\mathcal{B}_1 \mathcal{B}_2.
 \end{equation}
 Note that $\mathcal{H}_0$ can be viewed as the Hamiltonian in real space of a finite lattice with OBC and with a single site per unit cell. Note also that $\mathcal{H}_0$ is a Toeplitz matrix, i.e. a matrix in which each descending diagonal from left to right is constant. It can be readily shown that the bulk energy dispersion curve (Bloch Hamiltonian) associated to $\mathcal{H}_0$ is precisely $Q(\beta)$ given by Eq.(10), and the bulk energy spectrum $E_{OBC}$ is obtained  from Eq.(10) as $\beta$ varies in $\tilde{C}_{\beta}$.   
 \subsection{Some specific models}
 Non-Hermitian extensions of the SSH model, considered in several recent works \cite{R3bis,R5,BB1,BB2,R13,R14,BB4,BB5,BB5quatris,BB6,BB7,BB8,R16,SCh1,SCh2,F1,F2,F3,F4,F5}, provide paradigmatic examples of non-Hermitian topological two-band systems. They are obtain from Eq.(5) for specific form of $d_{x,y,z}(k)$. These models, originally introduced mostly as theoretical models, are nowadays being experimentally accessible with synthetic topological matter  using photonic structures and topolectrical circuits \cite{SCh2,F2,F3,Ruff0,Ruff3}. Other platforms, such as mechanical, acoustic, or other metamaterial settings, are also promising laboratory tools to physically implement non-Hermitian SSH models. In particular, the first experimental observation of the bulk boundary correspondence breakdown owing to NHSE in a SSH model with asymmetric hopping amplitudes has been very recently reported in non-reciprocal topolectrical circuits \cite{Ruff3}. Such experimental advances motivate us to focus our analysis to four non-Hermitian SSH models. Such models, already introduced in the recent literature,  are schematically shown in Figs.1(a) and (b) and capture most of the properties of non-Hermitian two-band systems, such as the presence (in models II,III,IV) or absence (in model I) of the NHSE, the existence of non-Bloch phase transitions (in models II and III), and the appearance of Bloch points (in model IV). For the sake of completeness, the main properties of such four SSH models are reviewed in Appendix A.\\ 
\begin{figure*}[htbp]
  \includegraphics[width=170mm]{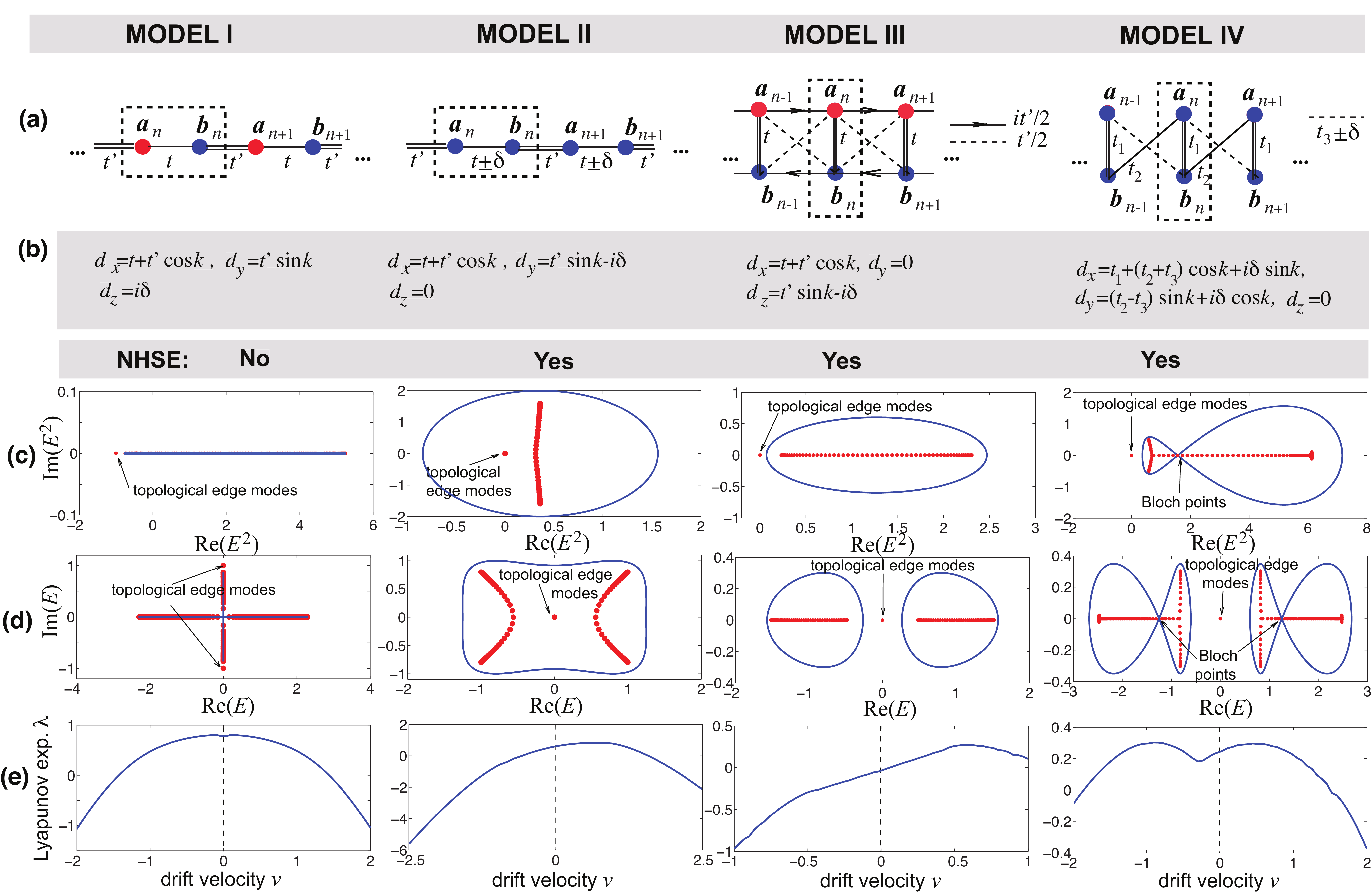}\\
   \caption{(color online) (a,b) Four examples of non-Hermitian SSH lattices, where non-Hermiticity is introduced by either onsite complex (gain/loss) energy (Models I and III) or by asymmetric hopping amplitudes (models II and IV). All models, with the exception of model I, show the NHSE. (c,d) Energy spectra  for systems with PBC (solid curves) and OBC (solid circles), showing both the behavior of $E^2$ [in (c)] and $E$ [in (d)]. Parameter values are: Model I, $t=1$, $t^{\prime}=1.5$, $\delta=1$; Model II, $t=0.6$, $t^{\prime}=1$, $\delta=1$; Model III, $t=0.6$, $t^{\prime}=1$, $\delta=0.3$; Model IV: $t_1=1$, $t_2=1.5$, $t_3=0.2$, $\delta=0.35$. In model I the PBC and OBC bulk energy spectra exactly overlap. (e) Numerically-computed behavior of the Lyapunov exponent $\lambda$ versus the drift velocity $v$. The largest value of the Lyapunov exponent is attained at $v=0$ in model I solely.}
\end{figure*} 
\section{Non-Hermitian skin effect and saddle point criterion}
The bulk energy spectra $E_{OBC}$ and $E_{PBC}$, corresponding to OBC and PBC boundary conditions, are distinct for systems displaying the NHSE, and a transition between them has been recently investigated in Ref.\cite{R14}. In this section we wish to establish a rather general criterion that relates saddle points of $Q(\beta)$ and the NHSE.\\
 As $\beta$ varies on the unit circle $C_{\beta}$, the energy spectrum $E_{PBC}=\pm \sqrt{Q(\beta)}$ describes a path in complex energy plane that can be  either a set of open arcs or one or more closed loops enclosing a non-vanishing area [solid curves in Figs.1(c) and (d)].  Empirically, it is found that in the former case (the energy spectrum $E_{PBC}$ is formed by a set of open arcs) the system does not show the NHSE and the bulk energy spectrum $E_{OBC}$ for OBC does coincide with   $E_{PBC}$ (model I in Fig.1); eventually, besides bulk states, the energy spectrum for OBC may include isolated points, corresponding to topological edge states.  In the latter case (the energy spectrum $E_{PBC}$ is composed by one or more closed loops enclosing a non-vanishing area) the system shows the NHSE and the bulk energy spectrum $E_{OBC}$ largely deviates from $E_{PBC}$, showing distinct symmetry breaking phase transitions (models II, III and IV in Fig.1). 
In the presence of the NHSE, the energy spectrum $E_{OBC}$ comprises a set of open arcs, which are located in the interior of their PBC loci \cite{R14}; see e.g. models II, III and IV in Fig.1. The bulk OBC spectrum is obtained from Eq.(9) with $\beta$ varying on the generalized Brillouin zone $\tilde{C}_{\beta}$. In the presence of the NHSE, the spectra $E_{PBC}$ and $E_{OBC}$ do not intersect or can intersect at isolated points, corresponding to so-called Bloch points \cite{BB5} separating bulk modes localized either at the left ($|\beta|<1$) or right  ($|\beta|>1$) edges of the system (see as an example model IV in Fig.1).\\  
A rather interesting property, that can be directly proven in the specific examples of Fig.1 and that we conjecture to be valid in rather general two-band systems (see Appendix B), is that the turning points of the open arcs forming the energy spectrum $E_{OBC}$ are attained at the values of $\beta$ on the generalized Brillouin zone $\tilde{C}_{\beta}$ that are the saddle points of $Q(\beta)$, i.e. where $(d Q/d \beta)=0$.  This means that the following criterion can be stated:\\ 
\\ {\it Saddle Point Criterion}.
If there exists at least one saddle point of $E^2=Q(\beta)$ that does not lie on the unit circle $C_{\beta}$, then the non-Hermitian Hamiltonian (5) shows the NHSE, and thus violates the Bloch bulk-boundary correspondence.\\
\\
The saddle point criterion gives a very simple sufficient condition for a non-Hermitian system to show the NHSE, however it is not strictly necessary;  an example of a non-Hermitian system with all saddle points on the unit circle that nevertheless shows the NHSE is presented in Appendix C. However, such cases are exceptional and occur under special singularity conditions, where the $E_{PBC}$ energy spectrum shows cusp singularities.

\section{Real-space wave packet dynamics and Lyapunov exponent} 
Let us consider the temporal dynamics in real space of a rather arbitrary wave packet {\it far from the edges} of a two-band non-Hermitian lattice. Our aim is to show that from the long-time behavior of the wave packet dynamics on the lattice one can (i) predict the existence of the NHSE and failure of Bloch bulk-boundary correspondence, and (ii) reveal the appearance of symmetry breaking phase transitions in the bulk OBC energy spectrum (i.e. non-Bloch phase transitions). To this aim, let
\begin{equation}
\left(
\begin{array}{c}
A \\
B
\end{array}
\right)_{\pm}=\frac{1}{\sqrt{2E_{\pm}(k) [E_{\pm}(k)-d_z(k)]}}
\left(
\begin{array}{c}
d_x(k)-id_y(k)\\
E_{\pm}(k)-d_z(k)
\end{array}
\right)
\end{equation}
be the (right) eigenvectors of $H(k)$ corresponding to the two lattice bands $E_{\pm}(k) = \pm \sqrt{Q(k)}$. In real space, the most general solution to the Sch\"odinger equation is given by a superposition of (extended) Bloch eigenfunctions
and reads 
\begin{equation}
\left(
\begin{array}{c}
a_n(t) \\
b_n(t)
\end{array}
\right)= \sum_{l= \pm} \int_{-\pi}^{\pi} dk F_{l}(k) \left(
\begin{array}{c}
A \\
B
\end{array}
\right)_{l}
\exp(ikn-i E_{l}t)
\end{equation}
where the spectral amplitudes $F_{\pm}(k)$ are determined by the initial excitation values $a_n(0)$, $b_n(0)$ on the lattice. We are interested to investigate the long-time behavior of the amplitude $\psi(t)=a_{n=vt}(t)$ (or similarly $\psi(t)=b_{n=vt}(t)$) along the space-time path $n=vt$, where $v$ is a drift velocity \cite{F6,F7}. The following properties can be proven:\\
(i)  The Lyapunov exponent
\begin{equation}
\lambda(v)= \lim_{t \rightarrow \infty} \frac{\log |\psi(t)|}{t}
\end{equation}
is bounded from above, namely $\lambda(v) \leq \lambda_{m}$ where $\lambda_{m}={\rm Im}(E_m)$ and $E_m$ is the energy of the PBC spectrum $E_{\pm}(k)$ with the largest imaginary part, taken at some value $k=k_0$ on ${C}_{\beta}$. Moreover, $\lambda(v)=\lambda_m$ for a drift velocity given by $v=v_m=( d {\rm Re} (E_{\pm}) / dk)_{k_0}$.\\
(ii) For a given drift velocity $v$, indicating by $k_s$ the (dominant) saddle point, satisfying the equation
\begin{equation}
\left( \frac{dE_{\pm}}{dk} \right)_{k_s}=v
\end{equation}
the Lyapunov exponent reads
\begin{equation}
\lambda(v)={\rm Im}(E_{\pm}(k_s))-v{\rm Im}(k_s).
\end{equation}
(iii) If $v_m \neq 0$, i.e. if the Lyapunov exponent $\lambda(v)$ does not exhibit its largest value at zero drift velocity, then the non-Hermitian model shows the NHSE. \\
\\
The last property states that a Lyapunov exponent exhibiting its largest value at a non-vanishing drift velocity is a clear signature of the existence of the NHSE, and thus of the breakdown of the Bloch bulk-boundary correspondence. A simple physical explanation of this result can be gained by considering the typical situation where the NHSE effect is observed, i.e. in the presence of an asymmetric hopping rate in the effective lattice described by dispersion curve $Q(\beta)$ \cite{nota}. As discussed in previous works \cite{R10,BB2,DW,palle}, in a lattice with OBC asymmetric hopping rates squeeze all bulk states towards one of the two edges \cite{R10,BB2}, while in a lattice with PBC a wave packet moving on the lattice is amplified or attenuated depending on its group velocity \cite{palle}, with the largest growth rate observed for a non-vanishing group velocity and the largest attenuation rate at the opposite group velocity. This means that, owing to the asymmetry of hopping amplitudes, the largest growth rate of a rather arbitrary initial excitation on the bulk of the lattice is observed along the space-time line $n=vt$ at the drift velocity $v$ that matches the group velocity with the largest growth rate.\\  
Let us now demonstrate the properties (i-iii) stated above. To this aim, let us consider the temporal evolution of the amplitude $\psi(t)=a_{n=vt}(t)$, along the space-time path $n=vt$, which is obtained from Eq.(22) after setting $n=vt$ and reads explicitly
\begin{eqnarray}
\psi(t) & = & \sum_{l= \pm} \int_{-\pi}^{\pi} dk G_l(k) \exp \left\{i[kv-E_{\l}(k)]t \right\} \\
& = & -i  \sum_{l= \pm} \int_{C_{\beta}} d\beta \beta^{v-1} G_l(\beta) \exp \left\{-iE_{\l}(\beta) t \right\} \nonumber
\end{eqnarray}
where we have set $G_l(k)=F_l(k)A_l(k)$ and $\beta=\exp(ik)$. Note that the temporal evolution of the amplitude $\psi(t)$ is fully determined by the interference of ordinary (extended) Bloch functions of the $E_{PBC}$ spectrum, while non-Bloch bulk states and $E_{OBC}$ spectrum do not seemingly play any role. To establish an upper bound for the Lyapunov exponent $\lambda(v)$, let us assume that the largest imaginary parts of $E_{\pm}(k)$, as $k$ spans the Brillouin zone $-\pi \leq k < \pi$, is attained at some value $k_0$, and let us indicate by $E_m$ the corresponding value of the most unstable band, either $E_{\pm}(k)$, at $k=k_0$. Then one has
\begin{eqnarray}
| \psi(t)|   & \leq  & \sum_{l= \pm} \int_{-\pi}^{\pi} dk \left| G_l(k) \exp \left\{i[kv-E_{\l}(k)]t \right\} \right| \nonumber \\
&   \leq & \exp [ {\rm Im}(E_m) t ] \sum_{l= \pm} \int_{-\pi}^{\pi} dk \left| G_l(k) \right| 
\end{eqnarray}
and thus
\begin{equation}
\frac{\log | \psi(t)| }{t} \leq {\rm Im}(E_m)+ \frac{1}{t} \log \left\{ \sum_{l= \pm} \int_{-\pi}^{\pi} dk \left| G_l(k) \right|  \right\}
\end{equation}
which yields, in the $t \rightarrow \infty$ limit, the following upper bound for $\lambda(v)$
\begin{equation}
\lambda(v) \leq {\rm Im}(E_m).
\end{equation}
The long-time asymptotic behavior of $\psi(t)$ can be determined rather generally by the
 steepest descent method \cite{F8}. This entails
analytic continuation of the functions $E_{\pm}(k)$ is the complex $k$
plane and, using the Cauchy theorem, the deformation of the
path of the integral along a suitable contour which crosses
the (dominant) saddle point $k_s$ of either $E_{+}(k)-kv$ or $E_{-}(k)-kv$
 in the complex plane, along the direction of the steepest descent \cite{F8}. 
 The dominant saddle point is the one with the largest imaginary part of $E_{\pm}(k)-kv$, corresponding to the largest 
 growth of $\psi(t)$ at long times. For a given drift velocity $v$, $k_s$ is obtained as one of the roots of the equation
 \begin{equation}
 \left( \frac{dE_{\pm}}{dk} \right)_{k_s}=v.
 \end{equation}
 \begin{figure*}
\includegraphics[width=17cm]{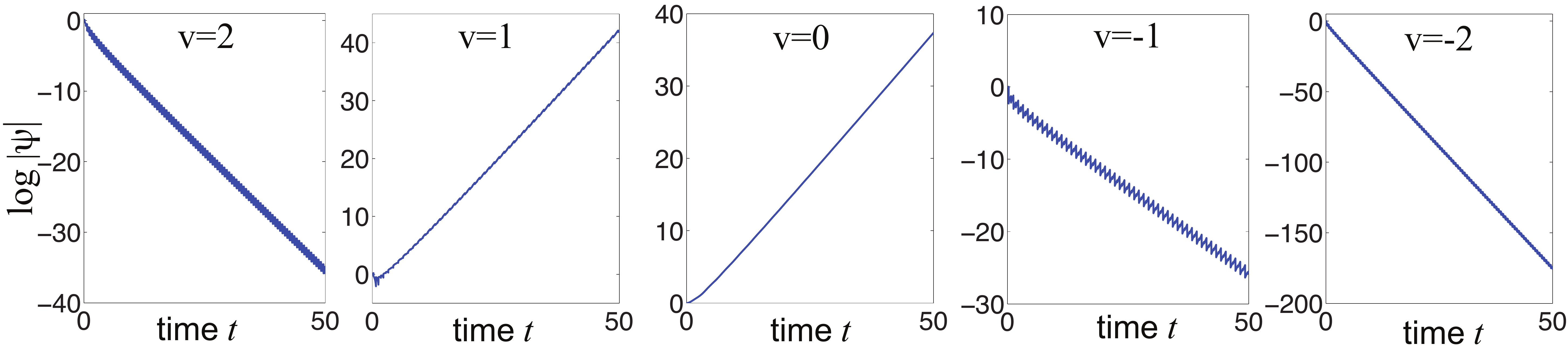}
\caption{(Color online) Numerically-computed behavior of the temporal evolution of $\log | \psi(t) |$, with $\psi(t)=a_{n=vt}(t)$, for the non-Hermitian SSH model II and for a few values of the drift velocity $v$. Parameter values are as in Fig.1 ($t=0.6$, $t^{\prime}=1$, $\delta=1$). Initial excitation of the lattice: $a_n(0)=b_n(0)=\delta_{n,0}$.}
\end{figure*}
 Assuming that there is one dominant saddle point of order $n \geq 2 $ at $k=k_s$, belonging for instance to the band $E_{+}$,
 the long-time asymptotic behavior of $\psi(t)$, as obtained by the steepest descend method, reads \cite{F8}
 \begin{eqnarray}
 \psi(t) & \sim  & \frac{G_+(k_s)}{\left| t \left( \frac{d^n E_+}{dk^n}\right)_{k_s} \right|^{1/n}}  \Gamma \left( \frac{1}{n} \right)\nonumber \\
 & \times & \exp \left[ \pm i \pi (n/2) +itvk_s-itE_{+}(k_s) \right]
 \end{eqnarray}
where $\Gamma$ is the Gamma function. From Eq.(31) it readily follows that
\begin{equation}
\lambda(v)=\lim_{t \rightarrow \infty} \frac{\log | \psi(t)|}{t}={\rm Im} \left( E_+(k_s)\right)-v {\rm Im}(k_s)
\end{equation}
which proves the property (ii) stated above.
We note that this result holds even if there are two (or more) dominant saddle points with the same growth rate, as it happens in systems with the symmetries $E \leftrightarrow -E$ and $E \leftrightarrow E^*$  of the energy spectrum. Let us now assume a drift velocity $v={\rm Re}(E_+(k_0)) \equiv v_m$. Then it can be readily shown that 
$E_{+}(k)-v_mk$ has a saddle point at $k=k_s=k_0$, i.e. $(dE_{+}/dk)_{k_0}=v_m$, $k_0$ being the (real) Bloch wave number where ${\rm Im}(E_{+}(k))$ on the PBC curve takes its largest value. For such a drift velocity, taking into account that ${\rm Im}(k_0)=0$  and letting $E_m=E_{+}(k_0)$, from Eq.(32) one has $\lambda(v_m)={\rm Im}(E_m)$. Such a result, together with Eq.(29), shows that the largest Lyapunov exponent is attained at the drift velocity $v=v_m$ and demonstrates the property (i) stated above.\\
The value $v=v_m$ of the drift velocity at which the Lyapunov exponent $\lambda(v)$ reaches its maximum value provides a clear signature of the existence (or not) of the NHSE for systems with OBC: apart from exceptional conditions, the NHSE arises whenever $v_m \neq 0$, as stated by property (iii). In fact, let us consider the value $\lambda(0)$ of the Lyapunov exponent at the zero drift velocity $v=0$. According to Eq.(30), the energy $E_+(k_s)$ of the dominant saddle point belongs to the OBC bulk spectrum, because $\beta_s=\exp(ik_s)$ is a saddle point of $Q(\beta)$ and any saddle point of $Q$ belongs to the bulk OBC energy spectrum (see Sec.III). If the system with OBC shows the NHSE, the energy $E(k_s)$ does not belong rather generally to the PBC energy spectrum,  i.e. $k_s \neq k_0$, unless the very exceptional case where there is a Bloch point which is also a saddle point (see Appendix C for more details). Hence from Eq.(32) with $v=0$ one has $\lambda(0)={\rm Im}(E_+(k_s))<{\rm Im}(E_+(k_0))=\lambda_m$, i.e. at the zero drift velocity $v=0$ the Lyapunov exponent does not reach its largest value. On the other hand, if the system with OBC does not show the NHSE, the bulk OBC and PBC energy spectra do coincide, and at $k=k_0$ (real)  the Bloch energy $E_+(k)$ has its dominant saddle point with ${\rm Re} (E_+(k_0))=0$. Hence in this case the largest value of the Lyapunov exponent is reached at the zero drift velocity. This proves property (iii) stated above.\\

The Lyapunov exponent in the long-time wave dynamics can be numerically computed by solving the coupled equations (1) and (2) with some given initial conditions, the result being insensitive to the specific initial condition. Typically, we initially prepared the lattice with excitation confined in one unit cell, namely $a_n(0)=b_n(0)=\delta_{n,0}$, and assumed a sufficiently long chain (comprising 150-500 unit cells) so that edge effects are avoided up to the maximum observation time $t_m$ (a propagation time $t_m \sim 10$ is usually sufficient to compute $\lambda$ with a good accuracy). The equations have been solved using an accurate variable-step fourth-order Runge-Kutta method. Some examples of temporal wave dynamics, showing the behavior of $\log |\psi(t)|$ versus time for the non-Hermitian SSH model II, are shown in Fig.2. The Lyapunov exponent $\lambda$ is obtained from the slope of the linear fit interpolation of the curves.  We checked that the obtained value of $\lambda$ is rather insensitive to the initial excitation condition; note that a propagation time of $ \sim 5-10$ is enough to estimate the linear fit with a good accuracy. Numerical results of Lyapunov exponent calculations for all other SSH models are summarized in Fig.1(e), clearly showing that the NHSE is associated to a non-vanishing value of $v_m$.\\ 
For models I, II and III we compared the numerical results of $\lambda(v)$ with the theoretical predictions based on Eq.(32). The results, shown in Fig.3, indicate an excellent agreement between the theoretical analysis and numerical simulations. For model I, the saddle points $k_s$, satisfying Eq.(24) for a given drift velocity $v$, are given by $\beta_s=\exp(ik_s)$, where $\beta_s$ is a root of the fourth-order algebraic equation
\begin{figure}
\includegraphics[width=8.5cm]{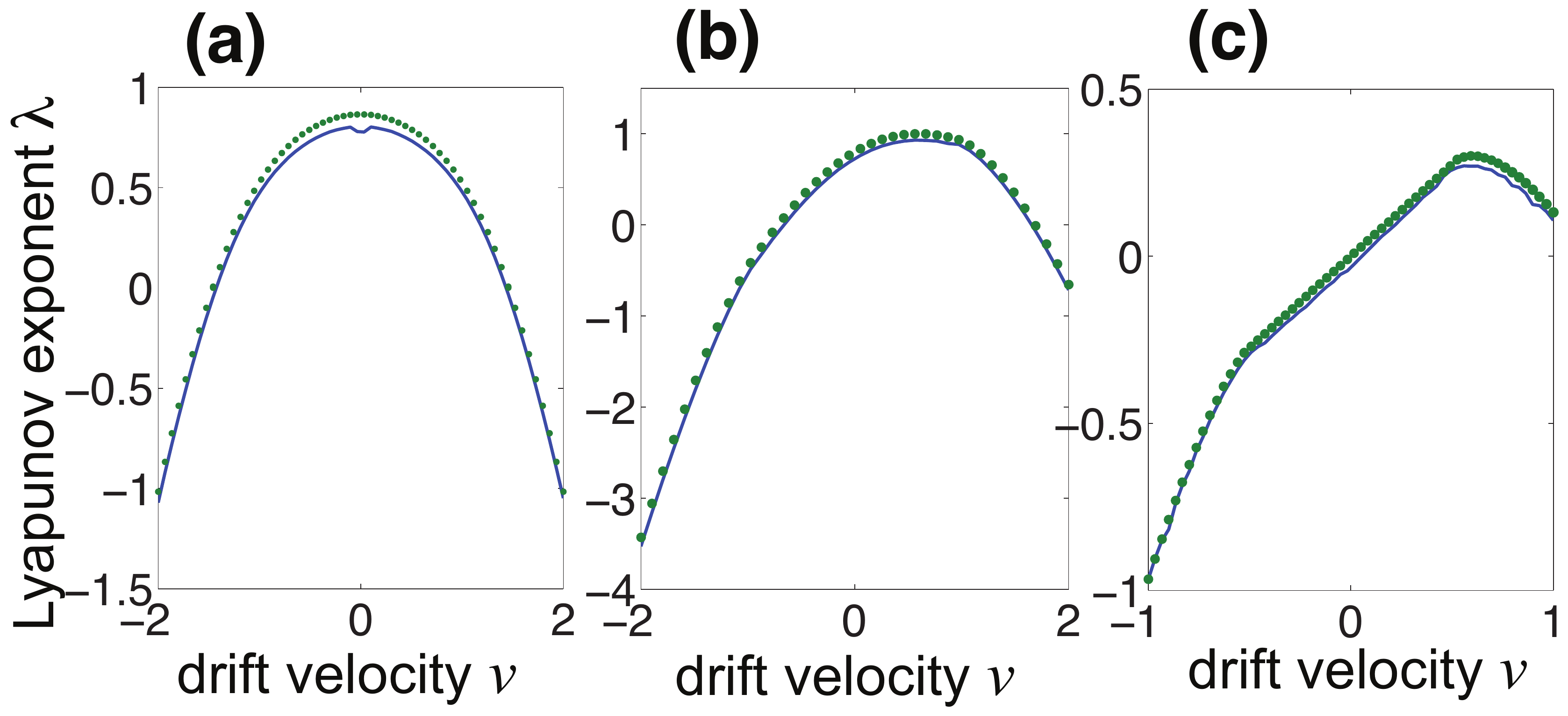}
\caption{(Color online) Behavior of the Lyapunov exponent $\lambda$ versus drift velocity $v$ for the non-Hermitian SHH model I [panel (a)], II [panel (b)] and III [panel (c)] for the same parameter values as in Fig.1. Solid curves refer to the numerical results obtained from the wave packet dynamics in real space, whereas dotted curves are the predictions of the steepest descent method.}
\end{figure}
\begin{equation}
\beta^4+c_1 \beta^3+c_2 \beta^2+c_3 \beta+ c_4=0
\end{equation}
with coefficients
\begin{eqnarray}
c_1 & = & c_3 = 4v^2 /(t t^{\prime}) \nonumber \\
c_2 & = & -2+4 v^2\frac{t^2+ t^{\prime 2}-\delta^2}{t^2 t^{\prime 2}} \nonumber \\
c_4 & = & 1.\nonumber 
\end{eqnarray}
The algebraic equation can be solved numerically, and the dominant saddle point, corresponding to the largest value of ${\rm Im}(E(k_s))-v {\rm Im}(k_s)$, is used in Eq.(32) to compute the Lyapunov exponent.  Likewise, for models II and III the saddle points $k_s$ are given by $\beta_s=\exp(ik_s)$, where $\beta_s$ is a root of the fourth-order algebraic equation (33) with coefficients
\begin{eqnarray}
c_1 & = &  4 v^2 \nonumber \\
c_2 & = & -2t^{\prime}(t+\delta)  +4 v^2\frac{t^2+t^{\prime 2}-\delta^2}{t^{\prime}(t-\delta)} \nonumber \\
c_3 & = & 4 v^2 \frac{t+\delta}{(t-\delta)}\nonumber \\
c_4 & = & \frac{t^{\prime} (t+\delta)^2}{(t-\delta)}. \nonumber
\end{eqnarray}
As a final comment, it should be noted that property (iii) above provides a sufficient condition for the existence of the NHSE, however it is satisfied in most cases of systems exhibiting the NHSE. Only in very special cases, corresponding to cusp singularities in the $E_{PBC}$ energy spectrum, one can observe $v_m=0$ in a system showing the NHSE, as discussed in the Appendix C.
\begin{figure}[htbp]
  \includegraphics[width=80mm]{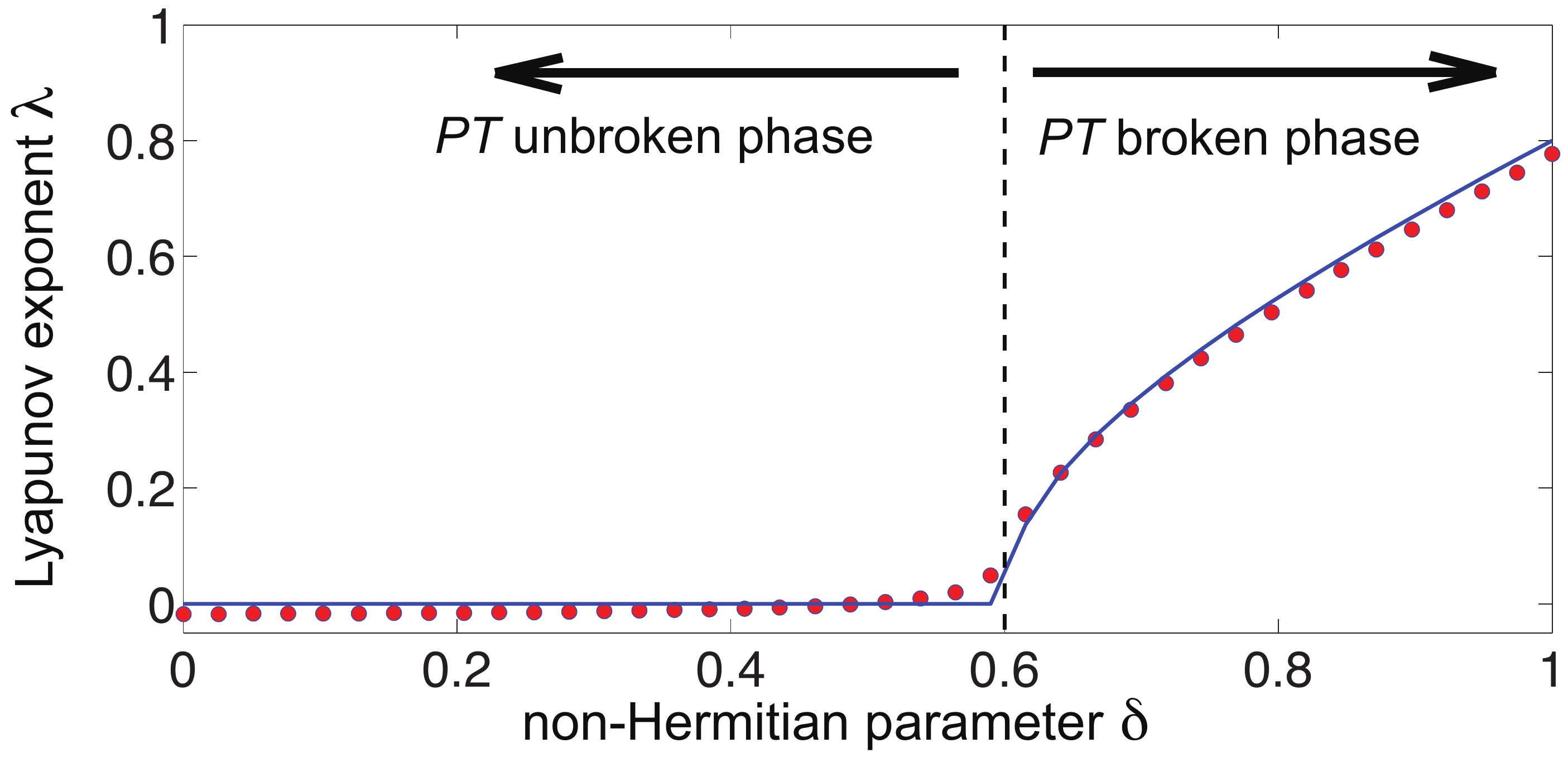}\\
   \caption{(color online) Behavior of the Lyapunov exponent $\lambda$ at zero drift velocity versus the non-Hermitian parameter $\delta$ for the SSH model III, revealing the $\mathcal{PT}$ symmetry breaking  phase transition of the OBC (non-Bloch) energy spectrum at $\delta=t$. Parameter values are $t=0.6$ and $t^{\prime}=1$. The solid curve refers to the theoretical value of $\lambda$ predicted by the steepest descent method [Eq.(36)], while the solid circles correspond to the Lyapunov exponent numerically-computed from wave packet dynamics in real space far from edges.}
\end{figure}

\section {Probing non-Bloch symmetry breaking phase transitions}
 The spectra $E_{PBC}$ and $E_{OBC}$ can undergo different symmetry breaking phase transitions as a non-Hermitian parameter in the system is varied. For example, let us consider model III shown in Fig.1, introduced by Lee in Ref.\cite{R3bis}. The Hamiltonian $H(k)$ has chiral ($\mathcal{S}$) and parity-time ($\mathcal{PT}$) symmetries, namely $\mathcal{S}H(k)=-H(k) \mathcal{S}$ and $\mathcal{P T}H(k)=H(-k)\mathcal{P T}$, where chiral, parity and time reversal operators defined by $\mathcal{S}=\sigma_y$, $\mathcal{P}=\sigma_x$ and $\mathcal{T}=\mathcal{K}$ ($\mathcal{K}$ is the element-wise complex conjugation). Note that the same symmetries can be introduced for the system with OBC. For a system with PBC, the $\mathcal{PT}$ symmetry is always in the broken phase for a non-vanishing value of the non-Hermitian (gain/loss) parameter $\delta$, while three different topological phases can be introduced, depending on the number of EPs that are encircled by the closed loop described by the gap vector $({\rm Re}(d_x(k)),{\rm Re}(d_z(k)))$ (see, for example, \cite{R8} and Appendix A). Conversely, for a system with OBC the $\mathcal{PT}$ phase remains unbroken for $|\delta|<t$ \cite{R3bis,BB2,BB5bis,R13}, i.e. the non-Bloch bulk energy spectrum $E_{OBC}$ undergoes a symmetry breaking phase transition which is not observed in the Bloch energy spectrum $E_{PBC}$. Our main result  here is that the non-Bloch symmetry breaking phase transition can be revealed from the wave packet dynamics on the lattice in real space far from any edge of the system.  To this aim, let us calculate the Lypaunov exponent $\lambda$ in the long-time dynamics for a drift velocity $v=0$. Since any saddle point $(d E_{\pm}/dk)_{k_s}=0$ belongs to the OBC energy spectrum, in the unbroken $\mathcal{PT}$ symmetry all saddle points are real and thus one has $\lambda=0$. On the other hand, in the broken $\mathcal{PT}$ phase the dominant saddle point corresponds to a positive imaginary part of the energy, leading to a non-vanishing value of the Lyapunov exponent. The  saddle points $k_s$ and corresponding energies $E_s=E_{\pm}(k_s)$ for model III can be calculated in a  closed form, and for $| \delta|>t$ they read explicitly
 \begin{eqnarray}
 k_s & = & -i \psi \pm \pi/2 \\
  E_s & = & \pm i \left( t \pm i t^{\prime} \sinh \psi \right) \sqrt{(\delta/t)^2-1}
 \end{eqnarray}
where $\psi$ is given by ${\rm tanh} \psi = t/ \delta$. From Eq.(25) with $v=0$ and Eq.(35), the steepest descent method thus predicts the following value of the Lyapunov exponent
\begin{equation}
\lambda= \left\{
\begin{array}{cc}
0 & |\delta| <t \\
\sqrt{\delta^2-t^2} & |\delta|>t.
\end{array}
\right.
\end{equation}
Therefore, measuring the Lyapunov exponent from real space wave packet dynamics far from any edge can reveal the non-Bloch symmetry breaking phase transition of a system with OBC. This is clearly shown in Fig.4, which depicts the numerically-computed Lyapunov exponent $\lambda$ versus $\delta$ in the temporal dynamics of a wave packet corresponding to initial unit cell excitation of the lattice ($a_n(0)=b_n(0)=\delta_{n,0}$). The numerically-computed value of the Lyapunov exponent turns out to be in very good agreement with the prediction (36) based on the steepest descent method.

\section{ Conclusion and Outlook}  A central principle in topological matter is that topological invariants of Bloch bands, detected by bulk dynamics in real space, can predict edge effects owing to the bulk-boundary correspondence. However, this main result can be violated in non-Hermitian systems. In such systems the bulk-boundary correspondence, formulated in terms of ordinary Bloch band invariants, can fail and the bulk energy spectrum for open boundaries can largely deviate from Bloch bands, showing distinct (non-Bloch) symmetry breaking phase transitions. The very distinct behavior of non-Hermitian systems under periodic and open boundary conditions, revealed by the non--Hermitian skin effect,  calls into question the usefulness of bulk dynamics to predict edge effects. In this work we have shown that, even though bulk dynamics in non-Hermitian systems is entirely described by Bloch band theory, the Lyapunov exponent in the  long-time dynamics is determined by the turning points of non-Bloch bands, which can reveal both non-Bloch symmetry breaking phase transitions and the existence  of the non-Hermitian skin effect.This means that, contrary to physical intuition, real-space wave packet dynamics, governed by Bloch-band theory, can reveal non-Bloch band features.
Our results are expected to stimulate further theoretical studies in a rapidly growing area of research, and could provide insights to experimental observation of non-Bloch phase transitions in photonic systems and topolectrical circuits, where non-Hermitian topological SSH models like the ones considered in this work can be physically realized \cite{SCh2,F2,F3,Ruff0,Ruff3}. There are some open questions ahead. For example, is the saddle-point method useful to predict non-Bloch band features in higher-dimensional models or in topological systems with synthetic dimensions? Since non-Bloch band features are basically determined by the saddle points of polynomials, can the non-Hermitian skin effect and violation of the Bloch bulk-boundary correspondence be linked to general properties of polynomials and number theory? 

\appendix
\section{Non-Hermitian SSH models}
Examples of non-Hermitian two-band systems include several extensions of the celebrated SSH model, which have been considered in several recent works (for a comprehensive review see \cite{R8}). Four models are schematically shown in Figs.1(a) and (b) and briefly reviewed here for the sake of completeness.\\
{\it Model I.} A first example of non-Hermitian SSH model, introduced in Ref.\cite{SCh1}, is obtained by assuming 
\begin{equation}
d_x=t+t^{\prime} \cos k \; , d_y=t^{\prime} \sin k \; , d_z=i \delta
\end{equation}
where $t$, $t^{\prime}$ are the intra- and inter-dimer hopping amplitudes, respectively, and $\delta$ is the complex onsite energy (alternating balanced gain and loss); see Fig.1(a).   
For this model one has
\begin{equation}
Q(\beta)=t^2+t^{\prime 2}-\delta^2+ t t^{\prime} \left( \beta+ \frac{1}{\beta} \right).
\end{equation}
As $\beta$ spans the unit circle $C_{\beta}$, $Q$ describes a segment on the real axis with extrema $Q_{-}=(t-t^{\prime})^2-\delta^2$ and $Q_+=(t+t^{\prime})^2-\delta^2$. The two turning points of the segment are attained at $\beta= \pm 1$, which are the saddle points of $Q(\beta)$. This model does not show the NHSE: the bulk energy spectrum of the lattice with OBC does coincide with the one with PBC and is given by $E_{\pm}= \pm \sqrt{Q}$. Additionally, for $t^{\prime}>t$, in the OBC system two topological edge states, at energies $\pm i \delta$ and localized at the left and right edges, are found.\\
\\
{\it Model II.} The second example, introduced in Ref.\cite{BB2}, deviates from the Hermitian SSH model because of asymmetric intra-dimer hopping  amplitudes. Such a model has been experimentally realized very recently in topolectrical circuits \cite{Ruff3}. The model is obtained by assuming 
\begin{equation}
d_x=t+t^{\prime} \cos k \; , d_y=t^{\prime} \sin k-i \delta \; , d_z=0
\end{equation}
where $t \pm \delta$ are the asymmetric intra-dimer hopping amplitudes whereas $t^{\prime}$ is the (Hermitian) inter-dimer hopping amplitude; see Fig.1(a). Like for the Hermitian SSH model, the Bloch Hamiltonian $H(k)$ has two important symmetries: chiral symmetry $\mathcal{S}=\sigma_z$ and time reversal symmetry $\mathcal{T}=\mathcal{K}$, i.e. $\sigma_z H(k)=-H(k) \sigma_z$ and $\mathcal{K}H(-k)=H(k) \mathcal{K}$, where $\mathcal{K}$ denotes the element-wise  complex conjugation. This means that the energy spectrum is invariant under the transformations $E \leftrightarrow -E$ and $E \leftrightarrow E^*$. For this model one has
\begin{equation}
Q(\beta)=t^2+t^{\prime 2}-\delta^2+\frac{t^{\prime}(t+ \delta)}{\beta}+{t^{\prime}(t- \delta)} \beta
\end{equation}
The energy spectrum $E^2$ for PBC is obtained by letting $\beta=\exp(ik)$ ($- \pi \leq k < \pi$) in Eq.(A4); it describes a closed loop (an ellipse) in complex $E^2$ plane, as shown in Fig.1(c). The square root generates the two bands $E_{\pm}(k)= \pm \sqrt{Q}$. Depending on how many EPs of $H(k)$, defined by $Q=0$, lie inside the contour described by $({\rm Re}(d_x),{\rm Re}(d_y))$, one obtains three different topological phases [1]: $|\delta|>|t+t^{\prime}|$ (none EP is enclosed in the contour), $|t-t^{\prime}|<|\delta|<|t+t^{\prime}|$ (one EP in enclosed in the contour), and $|\delta |<|t-t^{\prime}|$ (two EPs are enclosed in the contour). The corresponding PBC energy spectra are shown in Fig.5.\\
The bulk energy spectrum of the system with OBC strongly differs from the Bloch bands owing to the NHSE.
To calculate the spectrum for OBC, let notice that, for $|\delta|<t$, one can write
\begin{equation}
Q(\beta)=t^2+t^{\prime 2}-\delta^2+ t^{\prime} \sqrt{t^2-\delta^2} \left(  \exp(- \psi) \beta+ \frac{\exp(\psi)}{\beta} \right)
\end{equation} 
with ${\rm tanh} \psi= \delta/t$, while for $|\delta|>t$, one can write
\begin{equation}
Q(\beta)=t^2+t^{\prime 2}-\delta^2+ t^{\prime} \sqrt{\delta^2-t^2} \left( - \exp(- \psi) \beta+ \frac{\exp(\psi)}{\beta} \right)
\end{equation} 
with ${\rm tanh} \psi=t/ \delta$. Clearly, the condition $Q(\beta_1)=Q(\beta_2)$ with $| \beta_1|=|\beta_2|$ can be satisfied by letting $\beta=\exp(\psi+i \theta)$, with $\theta$ real varying in the range $(-\pi,\pi)$. 
This means that the generalized Brillouin zone $\tilde{C}_{\beta}$ is the circle of radius $\exp(\psi)$, while the energy spectrum for OBC reads explicitly:
\begin{equation}
E^2=t^2+t^{\prime 2}-\delta^2+ 2 t^{\prime} \sqrt{t^2-\delta^2}  \cos \theta
\end{equation}
for $|\delta|<t$, and 
\begin{equation}
E^2=t^2+t^{\prime 2}-\delta^2- 2 i t^{\prime} \sqrt{\delta^2-t^2}  \sin \theta
\end{equation}
for $|\delta|>t$, with $-\pi \leq \theta < \pi$. Equations (A7) and (A8) indicate that the OBC bulk spectrum remains real for $|\delta |< t$, while it becomes complex for $|\delta |> t$, corresponding to a $\mathcal{T}$ symmetry breaking phase transition; see Fig.5. At $|\delta|=t$, corresponding to unidirectional intra-dimer hopping, one obtains two high-order EPs for the matrix $\mathcal{H}$ in real space [Eq.(17)], with all bulk energies collapsing at the two EP energies  $E=\pm t^{\prime}$. It can be readily shown by a direct calculation that, for both $|\delta|<t$ and $|\delta|> t$, the turning points of the OBC bulk energy spectrum are the saddle points of $Q(\beta)$. The parameter range for the existence of topological edge states is derived in Ref.\cite{BB2}, where a non-Bloch bulk-boundary correspondence is established. Finally, from Fig.5 it follows that the  symmetry breaking phase transition of the non-Bloch energy spectrum for a system with OBC is unrelated to the three topological phases of a system with PBC.\\
 \begin{figure*}
\includegraphics[width=17cm]{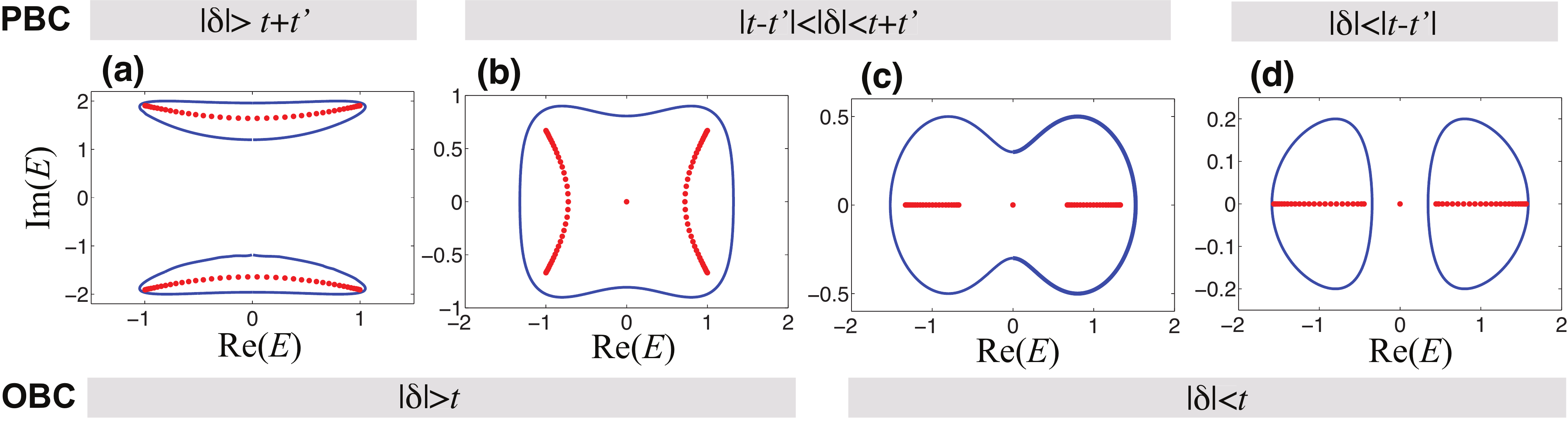}
\caption{(Color online) Behavior of energy spectra $E_{PBC}$ (solid curves) and $E_{OBC}$ (solid circles) in non-Hermitian SSH models II/III for parameter values $t=0.6$, $t^{\prime}=1$, and for (a) $\delta=2$, (b) $\delta=0.9$, (c) $\delta=0.5$, and (d) $\delta=0.2$. While in systems with PBC there are three different topological phases (upper row) and the energy spectrum is always complex for $\delta \neq 0$, in systems with OBC there are two distinct bulk phases (lower row) and a non-Bloch $\mathcal{T}$/$\mathcal{PT}$ symmetry breaking phase transition occurs at $| \delta |=t$.}
\end{figure*}
{\it Model III.} The third example of non-Hermitian SSH model, introduced in Ref.\cite{R3bis} and considered in several subsequent papers (see e.g.\cite{BB1,R13}), is obtained by assuming 
\begin{equation}
d_x=t+t^{\prime} \cos k \; , d_y=0 \; , \; d_z=t^{\prime} \sin k-i \delta .
\end{equation}
The real-space realization of this model is shown in Fig.1(a). Clearly, this model yields the same form for $Q(\beta)$ [Eq.(A4)] as model II. In fact, models II and III are basically equivalent and are obtained one another by exchanging $d_y$ and $d_z$; in real space the corresponding coupled-equations (1) and (2) are obtained one another after a unitary transformation (rotation) of the amplitudes $a_n$ and $b_n$. The Hamiltonian $H(k)$ has chiral (sublattice) and parity-time ($\mathcal{PT}$) symmetries, i.e. $\mathcal{S}H(k)=-H(k) \mathcal{S}$ and $\mathcal{P T}H(k)=H(-k)\mathcal{P T}$ with chiral, parity and time reversal operators defined by $\mathcal{S}=\sigma_y$, $\mathcal{P}=\sigma_x$ and $\mathcal{T}=\mathcal{K}$ ($\mathcal{K}$ is the element-wise complex conjugation). Note that the same symmetries can be introduced for the system with OBC, i.e. for the Hamiltonian $\mathcal{H}$. In fact, the explicit form of $\mathcal{H}$ is given by Eq.(17) with
\begin{equation}
\mathcal{A}=
\left(
\begin{array}{ccccccc}
-i \delta & -i t^{\prime}/2 & 0  & ... & 0 & 0  & 0\\
i t^{\prime}/2 & -i \delta & -i t^{\prime}/2  & ... & 0 & 0 & 0 \\
0 & i t^{\prime}/2 & - i \delta & ... & 0 & 0 & 0 \\
... & ... & ... & ... & ... & ... & ... \\
0 & 0 & 0 & ... & i t^{\prime}/2 & -i \delta & -i t^{\prime}/2 \\
0 & 0 & 0 & ... & 0 & i t^{\prime}/2 & -i \delta
\end{array}
\right)
\end{equation}
and
\begin{equation}
\mathcal{B}_1=\mathcal{B}_2=
\left(
\begin{array}{ccccccc}
t & t^{\prime}/2 & 0  & ... & 0 & 0  & 0\\
 t^{\prime}/2 & t &  t^{\prime}/2  & ... & 0 & 0 & 0 \\
0 & t^{\prime}/2 & t & ... & 0 & 0 & 0 \\
... & ... & ... & ... & ... & ... & ... \\
0 & 0 & 0 & ... &  t^{\prime}/2 & t & t^{\prime}/2 \\
0 & 0 & 0 & ... & 0 & t^{\prime}/2 & t 
\end{array}
\right)
\end{equation}
It then readily follows that $\mathcal{H}$ has chiral ($\mathcal{S}$) and parity-time ($\mathcal{PT}$) symmetries, i.e. $\mathcal{S} \mathcal{H}=-\mathcal{H} \mathcal{S}$ and $\mathcal{PT} \mathcal{H}=\mathcal{H} \mathcal{PT}$, with
chiral, parity and time-reversal operators defined by
\begin{equation}
\mathcal{S} \equiv i \left(\begin{array}{c|c} 0 & - \mathcal{I} \\\hline \mathcal{I} & 0 \end{array}\right) \; , \; \mathcal{P} \equiv  \left(\begin{array}{c|c} 0 &  \mathcal{I} \\\hline \mathcal{I} & 0 \end{array}\right) \; , \; \mathcal{T}= \mathcal{K}
\end{equation}
and where $\mathcal{I}$ is the $N \times N$ identity matrix.\\
The system is Hermitian in the limit $\delta=0$. As $|\delta|$ is increased above zero, like for model II one can distinguish three different topological phases of the PBC energy spectrum, depending on the number of EPs that are enclosed in the  loop described by $( {\rm{ Re}} (d_x), {\rm {Re}} (d_z))$ \cite{R8}.  $\mathcal{PT}$ symmetry is immediately broken in systems with PBC, while it remains unbroken in systems with OBC until $|\delta|$ reaches the symmetry breaking threshold $|\delta|=t$. This result indicates that, like for model II,  the symmetry breaking phase transition observed in systems with OBC in unrelated to the phases of systems with PBC.\\
\\
{\it Model IV.} The last example of non-Hermitian SSH model, introduced in Ref.\cite{BB5}, corresponds to the choice
\begin{eqnarray}
d_x(k) & = & t_1+(t_2+t_3) \cos k + i \delta \sin k \nonumber \\
d_y(k)&=&(t_2-t_3) \sin k +i \delta \cos k \\
d_z(k) & = & 0. \nonumber
\end{eqnarray}
The function $Q(\beta)$ for this model reads explicitly
\begin{equation}
Q(\beta)=\frac{(t_2 \beta^2+t_1 \beta+t_3-\delta)[(t_3+\delta) \beta^2+t_1 \beta + t_2]}{\beta^2}.
\end{equation}
Note that there are four saddle points of $Q(\beta)$, because the equation $(d Q / d \beta)=0$ is a quartic equation in $\beta$.
The Hamiltonian $H(k)$ has chiral ($\mathcal{S}= \sigma_z$) and time-reversal ($\mathcal{T}=\mathcal{K}$) symmetries. The system shows the NHSE, and the OBC bulk spectrum deviates from the PBC spectrum [see Fig.1(c) and (d)]. As the non-Hermitian parameter $| \delta |$ is increased above zero, the OBC energy spectrum shows a finite-threshold $\mathcal{T}$ symmetry breaking phase transition. On the other hand, the PBC energy spectrum is always in the broken $\mathcal{T}$ phase for a non-vanishing value of $\delta$. Interestingly, the generalized Brillouin zone $\tilde{C}_{\beta}$ can intersect the unit circle $C_{\beta}$ in two points, corresponding to so-called Bloch points \cite{BB5} and crossing of $E_{PBC}$ and $E_{OBC}$ spectral curves. The Bloch points are indicated by the arrows in Figs.1(c) and (d) of the main text. At these points, the bulk modes for OBC are extended, rather than being squeezed at the left or right edges \cite{BB5}.  
\section{ Bulk energy spectrum and saddle points in systems with OBC}
The bulk energy spectra $E_{OBC}$ and $E_{PBC}$, corresponding to OBC and PBC boundary conditions, are distinct for systems displaying the NHSE.
Empirically, it appears that the  $E_{PBC}$ spectrum describes one or more closed loops in complex energy plane, while $E_{OBC}$
comprises one or more open arcs internal to the PBC loops [see e.g. Figs.1(c) and (d)]. The transition from PBC to OBC energy spectra, based on an imaginary 
flux threading argument and showing the trajectories of PBC-OBC spectral 
flows, has been investigated in Ref.\cite{R14}. Here we disclose  a connection between turning points of the OBC spectral arcs and saddle points of $Q(\beta)$, which is essential to establish non-Bloch band features from bulk probing of wave dynamics in real space.\\
The first simple property is that any turning point of the open arcs, describing the OBC spectrum, is a saddle point of $Q(\beta)$. In fact, let $\beta$ a point on the generalized Brillouin zone
$\tilde{C}_{\beta}$ around $\beta_s \neq 0$, such that $Q(\beta_s)=E^2_V$ and $E_V$ is a turning point of the OBC spectrum. Since $Q$ is an analytic function of $\beta$ at around $\beta_s$, for $\beta \sim \beta_s$ one can write $Q(\beta) \simeq E_V^2+ (dQ/d \beta)_{\beta_s}(\beta-\beta_s)+...$. The condition that $E_V$ is a turning point clearly implies  that $(d Q / d \beta)_{\beta_s}=0$, i.e. $\beta_s$ is a saddle point of $Q(\beta)$. On the other hand, any saddle point of $Q(\beta)$ belongs to $\tilde{C}_{\beta}$. After writing $Q(\beta)= (q_0 \beta^M+ q_1 \beta^{M-1}+...+q_M)/ \beta^N$, with $q_{0,M} \neq 0$ and $N,M$ non-negative integers, the number of saddle points of $Q(\beta)$ is $M$. For the models I,II,III and IV described in Appendix A and shown in Fig.1, one has $M=2,2,2,4$, respectively. The number of turning points of $E_{OBC}^2$ in the four models is $2,2,2,4$ [see Fig.1(c)], and any saddle point $\beta_s$ of $Q$ belongs to the generalized Brillouin zone $\tilde{C}_{\beta}$. A non-Bloch phase transition, i.e. a phase transition of the bulk spectrum $E_{OBC}$, corresponds to the coalescence of saddle points. For example, in the model III discussed in Appendix A the function $Q(\beta)$ [Eq.(A4)] has two saddle points at $\beta_s= \pm \sqrt{(t+\delta)/(t-\delta)}$; the $\mathcal{PT}$ symmetry breaking phase transition observed at $|\delta|=t$ corresponds to the coalescence of the two saddle points.  We conjecture that such properties, checked for the four specific models, are rather general ones, in particular any saddle point $\beta_s$ of $Q(\beta)$ belongs to $\tilde{C}_{\beta}$. Albeit we are not able to provide a rigorous mathematical proof that the spectrum $E_{OBC}$ is composed by open arcs and this remains an empirical result \cite{R14}, we can show that at any saddle point $\beta_s$ of $Q$ the energy $E(\beta_s)$ belongs to $E_{OBC}$. In fact, let $\beta=\beta_s$ be a saddle point of $Q(\beta)$, with $\beta_s \neq 0$. Since $Q(\beta)$ is analytic at around $\beta_s$, for $\beta$ close to $\beta_s$ one can write
\begin{equation}
Q(\beta) \simeq Q(\beta_s)+\alpha (\beta-\beta_s)^n
\end{equation}
where $\alpha \neq 0$ and $n \geq 2$ is an integer.  Equation (B1) can be solved for $\beta$, yielding $n$ distinct branches
\begin{equation}
\beta=\beta_s \left\{ 1+ \left( \frac{Q-Q_s}{\alpha \beta_s^n} \right)^{1/n} \right\}
\end{equation}
where we have set $Q_s=Q(\beta_s)$. Let us now vary $Q$ around $Q_s$ by letting 
\begin{equation}
Q=Q_s+ \epsilon^n \alpha \beta_s^n
 \exp(i \phi)
 \end{equation}
 where $\epsilon \geq 0$ is a real parameter and $\phi$ a real phase, to be determined.  Substitution of the Ansatz (B3) into Eq.(B2) yields the following $n$ branches for $\beta=\beta_l$
 \begin{equation}
 \beta_l=\beta_s \left[ 1+ \epsilon \exp \left( i \frac{\phi + 2 l \pi}{n}\right) \right] \equiv \beta_s X_l
 \end{equation}
  \begin{figure}
\includegraphics[width=8.5cm]{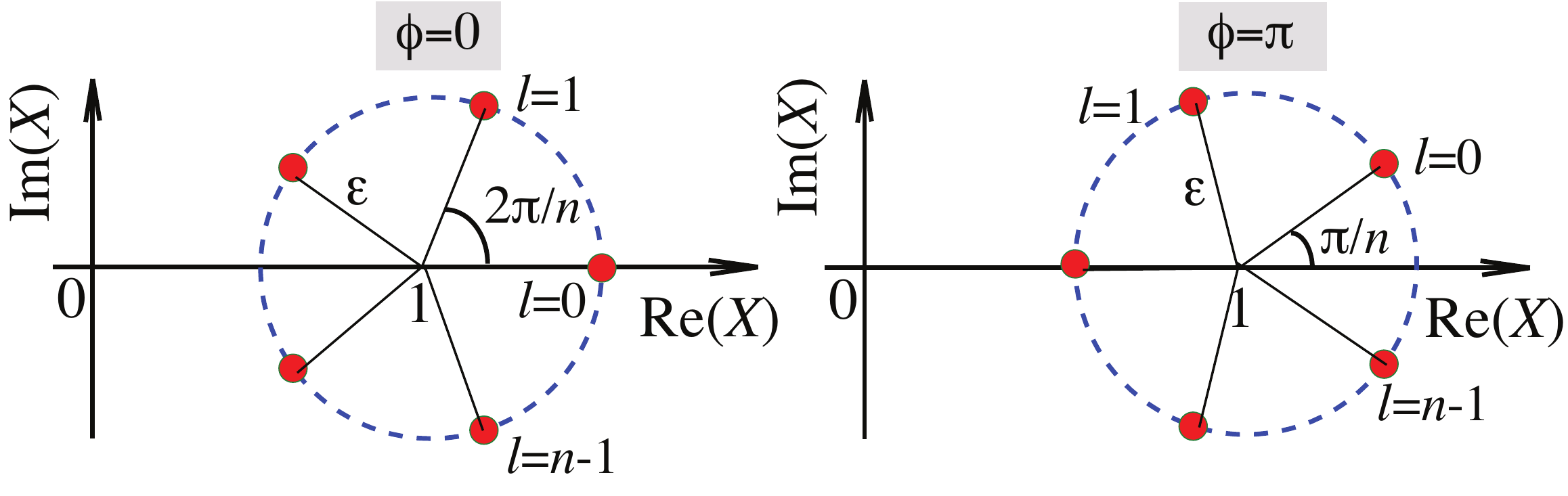}
\caption{(Color online) Loci in complex plane (solid circles) of the vectors $X_l=1+\epsilon \exp [i ( \phi + 2 l \pi)/n]$ ($l=0,1,2,...,n-1$)  for the two values of the phase $\phi=0$ (left panel) and $\phi=\pi$ (right panel).}
\end{figure}
 ($l=0,1,2,...,n-1$), which correspond to the same value of $Q$ given by Eq.(B3).  Clearly, by letting either $\phi=0$ or $\phi=\pi$, it readily follows from Eq.(B4) that one can find couples of values of $\beta$ on the distinct branches, say $\beta_{l_1}$ and $\beta_{l_2}$, such that $|\beta_{l_1}|=|\beta_{l_2}|$ (see the geometric construction of Fig.6). This means that we can always find two distinct values of $\beta$, $\beta_{l_1}$ and $\beta_{l_2}$, parametrized by $\epsilon$, such that  $|\beta_{l_1}|=|\beta_{l_2}|$ and $Q(\beta_{l_1})=Q(\beta_{l_2})$. Hence $\beta_{l_1}$ and $\beta_{l_2}$ are likely to belong to $\tilde{C}_{\beta}$. Interestingly, as the saddle (turning) point energy $Q_s$ is approached by letting $\epsilon \rightarrow 0$, one has 
 $\beta_{l_1} \rightarrow \beta_{l_2} \rightarrow \beta_s$, i.e. at the saddle point the two complex $\beta$ parameters coalesce. 
 The above argument, however, does not prove that near the saddle point  $\beta_s$ the curve $E_{OBC}^2$ describes an open arc with a turning point at $Q_s$; for that, one should exclude that $Q_s$ is a cusp. While saddle points at cusp singularities can be observed for $E_{PBC}^2$ (an example is discussed below and in Appendix C), in the models we considered and as in Refs.\cite{BB2,R14} we could not find saddle point cusps in $E_{OBC}^2$ spectra.\\
 In models showing the NHSE, the saddle points $\beta_s$ of $Q(\beta)$ do not belong rather generally to ${C}_{\beta}$, however in some special cases it might happen that a saddle point $\beta_s$ is also a Bloch point, i.e. $|\beta_s|=1$. An example of such an exception occurs in model IV. For the special choiche of parameters $t_2= \pm t_1/2$, the PBC energy spectrum shows a cusp at $\beta=\mp 1$,  which is a Bloch point and a saddle point of second order for $Q(\beta)$ (Fig.7). Another example of a system showing the NHSE, where all saddle points are also Bloch points and cusp singularities for $E_{PBC}$, is presented in Appendix C.\\
\begin{figure}
\includegraphics[width=8.5cm]{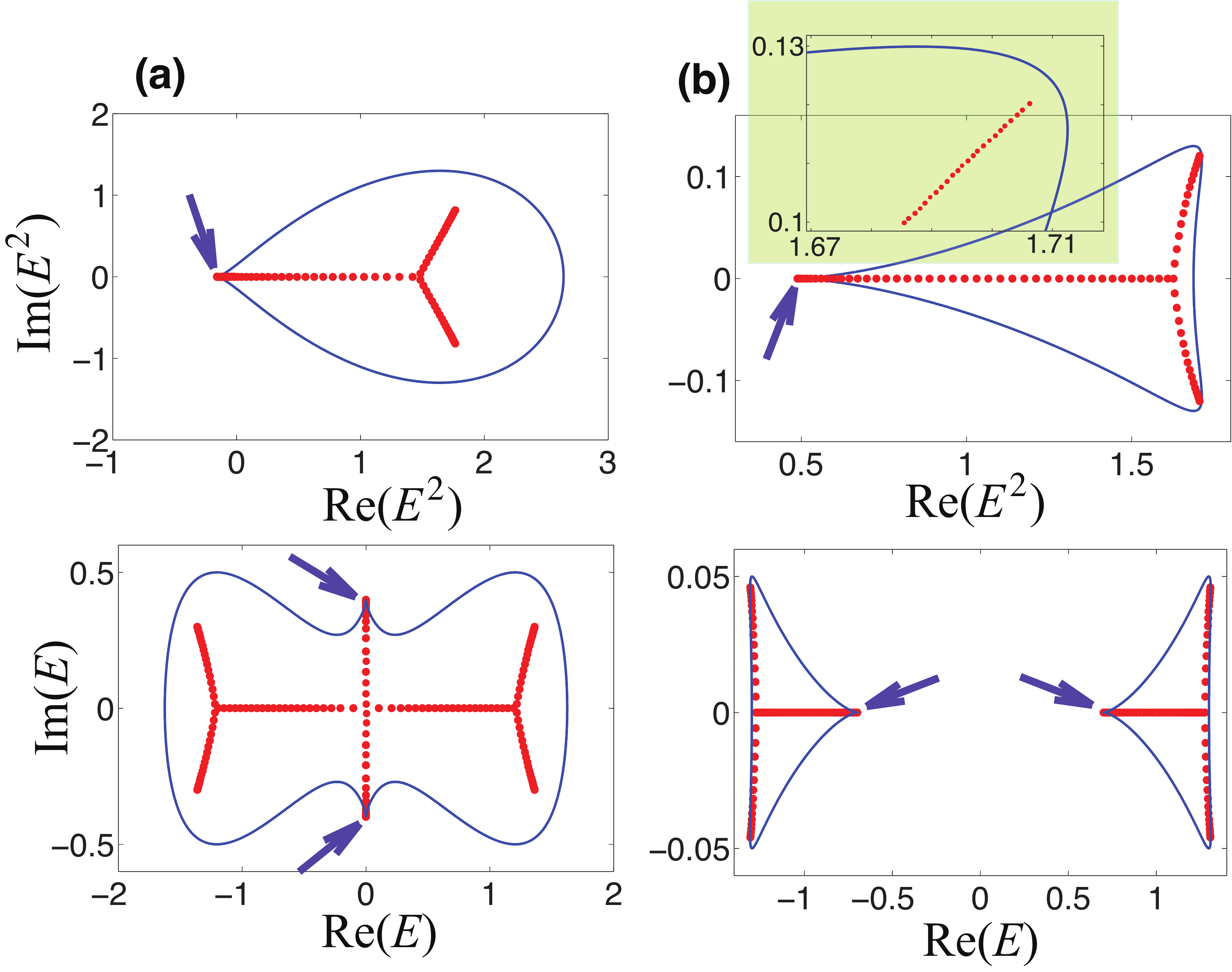}
\caption{(Color online) Behavior of the energy spectra ($Q=E^2$, upper panels, and $E$, lower panels) for the non-Hermitian SSH model IV corresponding to PBC (solid lines) and OBC (solid circles). Parameter values are: $t_1=1$, $t_2=0.5$, $t_3=0.2$, $\delta=0.5$ in (a), and $t_1=1$, $t_2=-0.5$, $t_3=0.2$, $\delta=0.05$ in (b). The arrows indicate a cusp in the $E_{PBC}^2$ / $E_{PBC}$ curves, corresponding to a saddle point of $Q(\beta)$ on the unit circle [$\beta=-1$ in (a), and $\beta=1$ in (b)]. The cusp is also a Bloch point, where the loci of $E_{PBC}$ and $E_{OBC}$ touch. The inset in the upper panel of (b) shows an enlargement of the energy curves at the top right region. Note that $E_{PBC}^2$ and $E_{OBC}^2$ do not touch in such a region.}
\end{figure}
\section{NHSE in systems with Bloch points and cusp singularities}
\begin{figure}
\includegraphics[width=8.5cm]{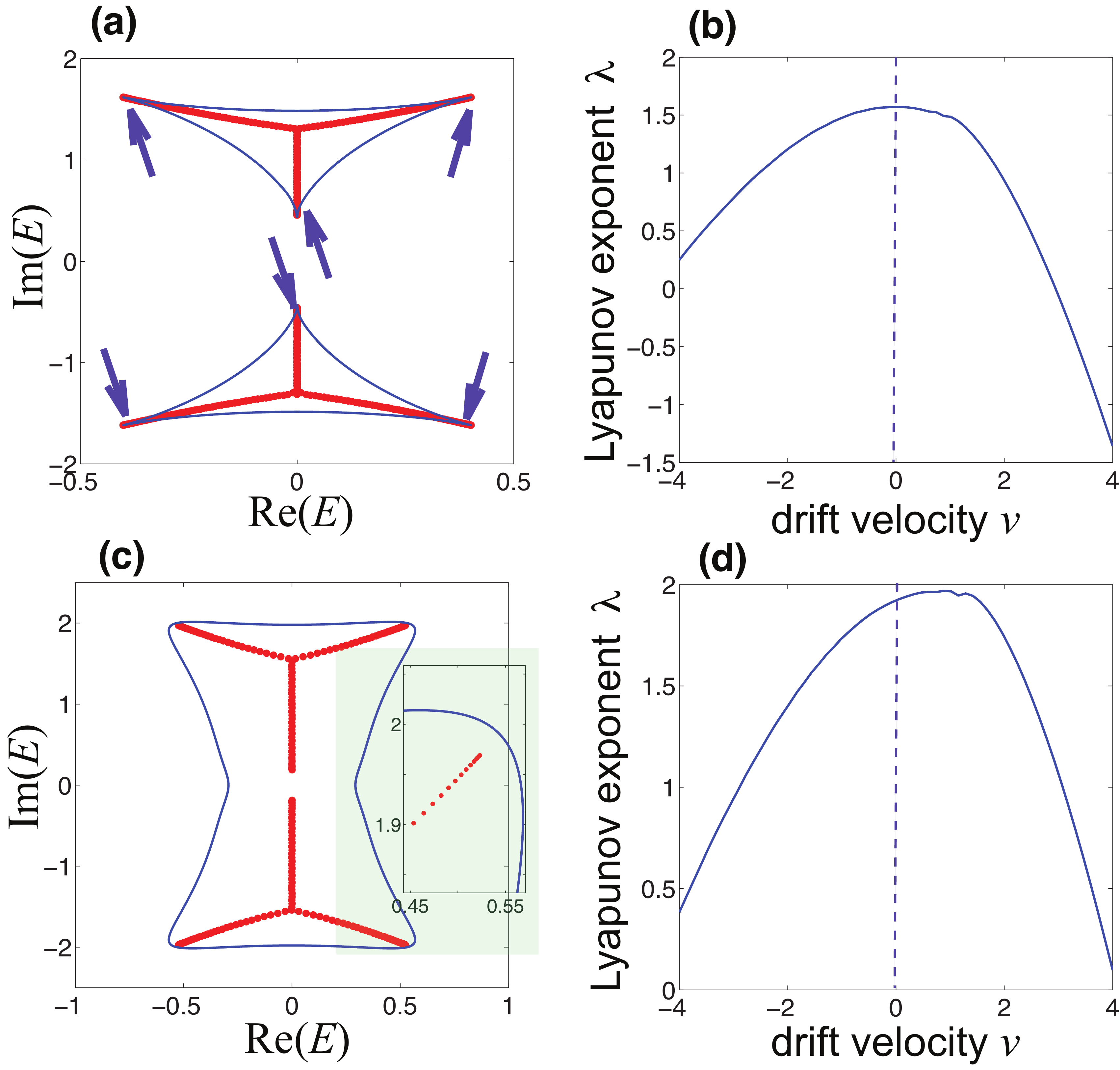}
\caption{(Color online) (a) Energy spectra $E_{OBC}$ for OBC (solid circles) and $E_{PBC}$ for PBC (solid lines) in the non-Hermitian model, defined by Eq.(C1), for parameter values $t=-1/2$ and $\delta=1$. Note that $E_{PBC}$ has six cusps, indicated by the arrows, which are Bloch points and also saddle points of $Q(\beta)$. (b) Corresponding behavior of the Lyapunov exponent $\lambda$ versus drift velocity $v$. (c,d) Same as (a),(b) but for parameter values $t=-1$ and $\delta=1$. The inset in (c) shows an enlargement of the energy spectrum of the lobe at the right top, indicating that the saddle points are not anymore Bloch points.}
\end{figure}
In non-Hermitian models showing the NHSE and exhibiting isolated Bloch points \cite{BB5}, it might exceptionally happen that the dominant saddle point of $Q(\beta)$ is also a Bloch point, or even that all saddle points of $Q(\beta)$ are Bloch points. In such special cases, the saddle-point criterion given in Sec.III is not satisfied, and the largest value of Lyapunov exponent is attained at $v=v_m=0$, even thought the system shows the NHSE. It should be emphasized that these are rather exceptional cases, usually observed when the $E_{PBC}$ energy spectrum shows cusps. A cusp singularity in the curve $E_{PBC}$ occurs whenever $\beta_s$ is a saddle point of $Q(\beta)$ on the unit circle, i.e. $|\beta_s|=1$, and $(dQ^2/d \beta^2)_{\beta_s} \neq 0$. Fortunately, even a small change of parameters in the system can shift the saddle point out of the unit circle and thus restore the validity of the saddle-point criterion and the condition $v_m \neq 0$ for a system to exhibit the NHSE.\\
To clarify the point, let us consider the two-band model with Bloch Hamitonian $H(k)$ defined by 
\begin{equation}
d_x=t \exp(ik)+\frac{\exp(-ik)}{\sqrt 2} , \; d_y=0, \; d_z=it \exp(ik)+i \delta
\end{equation}
which depends on the two real parameters $t$ and $\delta$. 
For this system, one has
\begin{equation}
Q(\beta)=\frac{-2 t \delta \beta^3+(\sqrt{2} t- \delta^2) \beta^2+1/2}{\beta^2}.
\end{equation}
There are three saddle points, which are the roots of the cubic equation
\begin{equation}
\beta^3+\frac{1}{2t \delta}=0
\end{equation}
i.e. 
\begin{equation}
\beta_s= \left(\frac{1}{2 t \delta} \right)^{1/3} \exp \left[  i \pi (2s+1)/ 3 \right]
\end{equation}
($s=0,1,2$). For the special values of parameters $2 t \delta =\pm 1$, one has $|\beta_s|=1$, i.e. all saddle points are also Bloch points, and the energies $\pm \sqrt{Q(\beta_s)}$ belong to both $E_{PBC}$ and $E_{OBC}$. Also, since $(d^2 Q / d \beta^2)_{\beta_s} \neq 0$, the $E_{PBC}$ energy spectrum shows cusp singularities at the energies of the Bloch points. Typical examples of energy spectra for the special condition $2 t \delta=-1$ are shown in Fig.8(a). The numerically-computed Lyapunov exponent $\lambda(v)$ is shown in Fig.8(b). Note that, since $E_{PBC}$ and $E_{OBC}$ do not coincide, the system  shows the NHSE. However, all saddle points of $Q(\beta)$ lie on the unit circle and, as shown in Fig.8(b), the largest value of the Lyapunov exponent is attained at $v=v_m=0$. However, as the special condition $2 t \delta =\pm 1$ is lifted, the saddle points are not anymore Bloch points, and the largest value of the Lyapunov exponent is reached at a non-vanishing drift velocity, as shown in Figs.8(c) and (d).


\begin{thebibliography}{31}
\bibitem{nota}
Effective asymmetric hopping rates are found in models II,III and IV looking at the Fourier coefficients of $Q(\beta)$, given by Eqs.(A4) and (A14) of Appendix A.
\end{thebibliography}
\end{document}